# Quantum bianisotropy in light-matter interaction

E. O. Kamenetskii

School of Electrical and Computer Engineering,
Ben Gurion University of the Negev, Beer Sheva, Israel

January 7, 2026

**Abstract**

Quantum bianisotropy and chirality are fundamental concepts in light-matter interaction that describe how materials with broken symmetries respond to electromagnetic fields at the level of macroscopic quantum electrodynamics. In quantum bianisotropy, magnetoelectric (ME) energy plays a critical role in mediating and enhancing light-matter interactions. This concept is essential for bridging the gap between classical electromagnetics (where bianisotropy often involves field non-locality) and quantum mechanics in metamaterials. The precise manipulation of a quantum emitter's properties at a subwavelength scale is due to near fields, which effectively function as a tunable environment. We show that the ME near field, interpreted as a structure combining the effect of bianisotropy/chirality with a quantum atmosphere, is a non-Maxwellian field with space-time symmetry breaking. Quantum ME fields arise from the dynamic modulation and topological coupling of magnetization and electric polarization within ME meta-atoms – specific subwavelength structural elements with magnetic and dielectric subsystems in magnetic insulators.

## I. INTRODUCTION

Quantum bianisotropy should be based on quantum ME meta-atoms to provide local ME coupling in a subwavelength domain, which is necessary for defining a quantum ME energy and exploring asymmetric momentum transfer from vacuum fluctuations (Feigel effect [1, 2]). Recent research discusses the relationship between ME energy, bianisotropy, and ME meta-atoms, suggesting that a small ferrite resonator can serve as a model for a building block of a metamaterial structure with such properties. The ME response in a ferrite disk model is characterized by the violation of parity ($P$) and time-reversal ($T$) symmetry, key aspects related to quantum bianisotropy. The properties of quantum meta-atoms are intrinsically linked to the localized near fields they can produce and interact with. Characterizing these evanescent field couplings is vital for understanding how a lattice of meta-atoms behaves collectively.

EM wave propagation is the process of coupling between polar (e.g., $E$-field) and axial (e.g., $H$-field) vectors in a *region the size of a wavelength*. In the dynamical regime of ME materials and ME samples, coupling between polar and axial vectors occurs in the *subwavelength region* of EM radiation. One of the basic questions about whether electromagnetism and magnetoelectricity *can coexist without an extension of Maxwell's theory* arises when we study EM wave scattering from subwavelength resonant objects. Currently, subwavelength resonators are considered as structural elements in *chiral and bianisotropic* metamaterials. It is assumed that these individual resonant scatterers behave as meta-atoms with strong electric and magnetic responses. There is a general consensus that these classical scatterers (modeled as *LC*-circuit elements) can be described by ME point–dipole interactions. In an analysis of ME interactions in bianisotropic metamaterials, simple electrodynamic models with dipolar (electric and magnetic) terms are used [3 – 5].

However, the main point is that there are *no near-field solutions of Maxwell's equations* with two sources, electric and magnetic currents, which are supposedly linked by electromagnetic forces in a *subwavelength (quasistatic)* region. The realization of *local coupling* between electric and magnetic dipoles must be associated with the violation of both spatial and temporal inversion symmetries. The "first-principle", "microscopic-scale" ME effect of a structure composed by "glued" pairs of electric and magnetic dipoles raise questions on the methods of *local probing* the dynamic ME parameters. The model of bianisotropic metamaterials (materials composed of *bianisotropic particles*) typically accounts for the ME response as a *far-field characteristic* resulting from the geometry and arrangement of the constituent subwavelength elements, as opposed to the *local*, intrinsic ME energy of interaction between separate, point-like electric and magnetic dipoles in the *near-field region*. In contrast, materials with *intrinsic* magnetoelectricity possess a *local ME energy density* due to specific microscopic or mesoscopic physical mechanisms, often comprising a violation of both spatial and time-reversal symmetries [6 – 8]. This involves the actual energy of interaction between electric and magnetic subsystems within the ME meta-atom [9 – 12], which can be *observed in the near-field*.

Chiral particles are considered as a class of bianisotropic particles characterized by a particular symmetry (space symmetry breaking) in their ME coupling. Within the framework of the dipole approximation, a chiral particle is represented as a pair of coupled electric and magnetic dipoles. Chiral objects can only be distinguished through interaction with other objects that have a different *3D arrangement* (opposite enantiomers). The difference also becomes apparent when they interact with such a chiral environment as circularly polarized light (CPL). The CPL technique is effectively used to distinguish enantiomers by considering the chiral response as a far-field characteristic. Chirality, like bianisotropy, is a form of nonlocal property. Non-local materials exhibit a response that depends on the field distribution over a spatial extent [13].

The utilization of plasmonic materials that could control local chiral light–matter interactions successfully facilitated chiral sensing into ultrasensitive detection. To enhance the subtle differences between enantiomers' interactions with light, various types of engineered chiral fields, called *superchiral fields*, have been proposed. These fields, being generated by plasmonic structures, are crucial for advanced sensing [14 – 16]. The superchiral fields described by the parameter of chirality *C* are referred to as inherently 3D near fields [17, 18]. It is stated that the superchiral fields are *near fields* because they create intense, *non-propagating* electromagnetic fields right at the surface, with spatial variations much smaller than a wavelength. However, the use of such artificially created chiral fields – superchiral fields – in an analysis of plasmonic structures reveals an obvious paradox. In [17], the "near fields" are created based on two *counterpropagating* circularly polarized waves with slightly different intensity. *Interference* of these propagating plane waves, giving enhanced chiral asymmetry on a 2D region is not actually a 3D near-field structure. Situation with plasmonic chiral meta-atoms is quite different from Tang and Cohen's model. Light coupled to electron oscillations on a metal surface, used in nanophotonics, are *slow EM waves*. The confinement ("trapping") occurs due to *evanescent fields* decaying perpendicular to the metal surface. The total distance *along the boundary* of a 2D-shape plasmonic sample is about a wavelength of the free-space EM wave. Therefore, we have *EM surface-wave resonance* along the perimeter. These are not the superchiral Tang and Cohen fields and, certainly, it is not a 3D confinement effect of chirality in a subwavelength region. Obviously, in an analysis of plasmonic chiral meta-atoms, Tang and Cohen's parameter of chirality *C* is used very formally. The generation of "superchiral" nodes observed in experiments [14 – 16] can be explained in terms of signal amplification arising from the well-known effect of surface plasmon amplification in systems fabricated with a metal substrate.

The Tang and Cohen parameter of chirality (optical chirality) [17], used to describe the interaction of light with chiral molecules, is proportional to the *imaginary part* of the product of the electric and magnetic fields. This time-even pseudoscalar measures the degree of chiral asymmetry in an electromagnetic field and is related to the spin angular momentum of light. On the other hand, the so-called parameter of magnetoelectricity [19] is proportional to the *real part* of the product of the electric and magnetic fields. This quantity is claimed to be a time-odd pseudoscalar and describes the effects related to the breaking of both parity and time-reversal symmetries. Similar to the chiral fields in [17], the "magnetoelectric" fields in [19] are specific fields created by two *counterpropagating* polarized waves which slightly different intensity. The effect should be observed due to the *interference* of these propagating plane waves on a 2D region. In [20], the idea to obtain the subwavelength structure with vacuum fields with the real product of the electric and magnetic fields was realized based on special plasmonic-vortex structure. However, *PT* properties of the fields in this structure are related to the parameter obtained in the plane wave propagation regime [19]. This cannot be considered as a local 3D-confined effect.

From Maxwell's electrodynamics, it is impossible to observe magnetoelectrically coupled electric and magnetic evanescent fields in the 3D vacuum space. Meta-atoms described in [17] can be referred as *chiral meta-atoms* while meta-atom described in [19] can be referred to as the Tellegen meta-atoms. Chiral/Tellegen meta-atoms are not subwavelength 3D open resonators. Both types of these meta-atoms are classical systems whose properties are due to their specially created structures with EM responses and are not quantum emitters like semiconductor quantum dots (QDs). In semiconductor QDs, the electron *effective mass* is generally different from the bulk material value and is a function of the dot's size, shape, and material composition. Electron energy levels are discrete and determined by quantum confinement, often modeled as a "particle-in-a-box" or harmonic oscillator potential. Fundamental differences between the near fields of classical chiral/Tellegen meta-atoms and quantum emitters arise from the inherent properties of their excited states and the nature of their light-matter interaction. The near fields of chiral/Tellegen meta-atoms are Maxwellian electromagnetic fields. Their interaction with incident light is described by constitutive relations, which, in fact, are *non-local bianisotropic* constitutive relations [13]. The near fields of semiconductor QDs originate from the *quantized* energy transitions. It is intrinsically linked to the quantum mechanical wavefunctions.

The quantum behavior of bianisotropic metamaterials should arise from quantum effects in ME meta-atoms. These quantum properties of ME meta-atoms include quantized energy states, quantum vacuum effects, and the ability to extract linear or angular momentum from quantum vacuum fluctuations.

## II. PROPAGATION OF ELECTROMAGNETIC WAVES IN BIANISOTROPIC MEDIA

The study of the quantum properties of electromagnetic EM wave propagation in bianisotropic media involves applying principles of canonical quantization to these complex materials. A bianisotropic medium is modeled by introducing two independent sets of three-dimensional harmonic oscillators to represent the material's electric and magnetic polarization fields, which are then coupled to the quantized EM fields [21, 22]. In the study of quantum properties of EM wave propagation in standard bianisotropic media, *ME energy* is typically not included. These materials, modeled in the far-field limit, are generally considered to be time-even effects related only to the violation of spatial inversion symmetry. This means that in this analysis, the ME effect is considered as the effect of nonlocality arising from the geometry of subwavelength structures, rather than an intrinsic, *local* ME coupling.

Being *not electromagnetic in nature*, ME energy density plays a crucial role in forming a topological structure of the fields. This can be well understood from classical analysis of EM wave propagation in bianisotropic media. By considering Poynting's theorem for a homogeneous material with electric, magnetic and ME characteristics, some important fundamental aspects of the relationship between electromagnetism and bianisotropy become clear.

Based on the initial assumption that in the dipole approximation, a bianisotropic electromagnetic material is a composition of coupled electric and magnetic dipoles, the constitutive relations are described by four tensors:

$$\vec{D}(\omega) = \vec{\vec{\varepsilon}}(\omega)\vec{E} + \vec{\vec{\xi}}(\omega)\vec{H}, \quad \vec{B}(\omega) = \vec{\vec{\zeta}}(\omega)\vec{E} + \vec{\vec{\mu}}(\omega)\vec{H}. \quad (1)$$

To derive the energy balance equation in temporary dispersive dielectric and magnetic media, one must use the regime of propagation of *quasi-monochromatic* EM waves [6]. For a ME medium, such quasi-monochromatic behavior was considered in Ref. [9]. The fields are expressed as $\vec{E} = \vec{E}_m(t,\vec{r}) \, e^{i(\omega t - \vec{k}\cdot\vec{r})}$ and $\vec{H} = \vec{H}_m(t,\vec{r}) \, e^{i(\omega t - \vec{k}\cdot\vec{r})}$, where complex amplitudes $\vec{E}_m(t,\vec{r})$ and $\vec{H}_m(t,\vec{r})$ are time and space smooth-fluctuation functions. It should be assumed [6, 9] that in the frequency regions of the transparency of the EM-wave propagation, the concepts of an internal-energy densities in alternative fields are introduced in the same sense as they are used in the *electrostatic*, *magnetostatic* and *magnetoelectrostatic* structures [23]. For the time-averaged stored energy $\langle W \rangle$ we have:

$$\langle W \rangle = \frac{1}{4}\left\{\frac{\partial\left(\omega \varepsilon_{ij}^h\right)}{\partial \omega}E_i^* E_j + \frac{\partial\left(\omega \mu_{ij}^h\right)}{\partial \omega}H_i^* H_j + \frac{\partial\left[\omega\left(\zeta_{ij}^h + \xi_{ij}^h\right)\right]}{\partial \omega}\left(H_i^* E_j\right)^h + \frac{\partial\left[\omega\left(\zeta_{ij}^{ah} - \xi_{ij}^{ah}\right)\right]}{\partial \omega}\left(H_i^* E_j\right)^{ah}\right\} \quad (2)$$

Superscripts *h* and *ah* in Eq. (2) denote, respectively, the Hermitian and anti-Hermitian parts of tensors of the second rank. For a lossless medium, tensors $\vec{\vec{\varepsilon}}$ and $\vec{\vec{\mu}}$ are Hermitian and $\vec{\vec{\xi}}^h = \vec{\vec{\zeta}}^h$, $\vec{\vec{\xi}}^{ah} = -\vec{\vec{\zeta}}^{ah}$. In a quasistatic limit, $\lambda \to \infty$, the parts of average stored energy in Eq. (2) correspond to potential energies in a *magnetoelectrostatic* structure. These are the electric-field energy density $W_E$ and the magnetic-field energy density $W_M$ expressed by the first two terms on the right-hand side of Eq. (2). The ME energy density $W_{ME}$ is expressed by the last two terms on the right-hand side of Eq. (2).

It is very important to note that the continuity equation for energy (Poynting's theorem) for ME medium is valid only when definite constraints are imposed to slowly time-varying amplitudes of the field components [9]:

$$E_{m_i}^*(t)\frac{\partial H_{m_j}(t)}{\partial t} = H_{m_j}(t)\frac{\partial E_{m_i}^*(t)}{\partial t}. \qquad (3)$$

The physical meaning of these constraints is that subwavelength fluctuations in the intensity of EM excitation in the ME medium occur *only at certain ratios of the complex amplitudes* of the electric and magnetic fields. The equation (3) can be transformed as follows:

$$\frac{E_{m_i}^*(t)}{H_{m_j}(t)} = \frac{\partial E_{m_i}^*(t)/\partial t}{\partial H_{m_j}(t)/\partial t} = \frac{dE_{m_i}^*(t)}{dH_{m_j}(t)}, \qquad (4)$$

where $dE_{m_i}^*$ and $dH_{m_j}$ are differentials of the corresponding fields. Eq. (4) implies that there exists a linear time-relation coupling between complex amplitudes of the fields. In a general form, we can write [10]

$$E_{m_i}(t) = T_{ij} H_{m_j}^*(t). \qquad (5)$$

The matrix $[T]$ is a field-polarization matrix. This is an invariant defined for a specific type of ME medium. Components of matrix $[T]$ are complex quantities. To find parameters of the polarization matrix $[T]$, one should solve an electromagnetic boundary problem with the known constitutive parameters of the electric and magnetic subsystems of the medium. When the parameters of matrix $[T]$ are found, an average energy density in a bianisotropic medium can be determined. In general, we have an integro-differential problem.

The derivation of the continuity equation for energy from symmetry principles is related to Noether's theorem. By virtue of Noether's theorem, the invariance of the action under translation in time, translation in space, and rotation, implies the existence of the conservation of energy, linear momentum, and angular momentum, respectively. Noether's theorem is used to investigate symmetries and related conserved quantities in Maxwell's equations [24]. With Noether's theorem, we can demonstrate electric–magnetic symmetry in ME electromagnetism. The constraints (3) – (5) are considered as certain symmetry conditions for the field structure. It becomes obvious that *without these constraints it is impossible to use the concept of ME energy* [9].

We know that for any EM process in a lossless non-ME medium, $|\langle W_E \rangle| = |\langle W_M \rangle|$. For a propagating monochromatic plane wave, the electric energy density and magnetic energy density are equal to each other at every instant of time. In the case of an EM resonator (such as, for example, a closed metal-wall cavity or a dielectric resonator), we have a quasistatic process when electrical energy is converted into magnetic energy and vice versa over a time period of the EM radiation. "Mediators" of these transformations are electric conduction currents and/or electric currents of polarization (displacement currents). Now, in electrodynamical ME effects, we have the *three parts* of the average stored energy, $\langle W_E \rangle$, $\langle W_M \rangle$, and $\langle W_{ME} \rangle$. In this case, it must be assumed that there is also a specific "mediator" associated with the ME effect, which mutually converts the electric and magnetic energies. In a *subwavelength region* where both the electric and magnetic fields exist, energy conversion between average stored energies $\langle W_E \rangle$, $\langle W_M \rangle$, and $\langle W_{ME} \rangle$ can occur through a certain *circular process of energy exchange.* This implies the presence of a *local power-flow circulation*, which can be carried out with *synchronous rotation* in time of both complex-amplitude vectors $E_{m_i}$ and $H_{m_j}$ [9, 10]. So, the process is accompanied by *power-flow vortices* in subwavelength domains of EM radiation. The power flows of the circulating adiabatic processes of the energy interchange are not exclusively the EM power flows. In structures where temporal inversion symmetry is broken, we can observe the right- and left-hand vortices.

Our analysis of the energy balance in bianisotropic metamaterials leads us to important conclusions. First, it should be noted that in these metamaterials, the *material and field structures are integrally linked and cannot be considered separately.* In this medium-field system, the intrinsic dynamics of the ME meta-atom determines the symmetry properties of the field (we call such a field the ME field). The ME field, in turn, determines the *intrinsic dynamics of the ME meta-atom*. The energy balance equation for EM waves propagating in the ME medium, showing the unique effect of power-flow circulations in local regions, reveals the fact that such a subwavelength circulation quantum process of energy should also occur in small ME resonators – ME meta-atoms. In this meta-atom, we have a ME "trion"– a localized (subwavelength) resonant excitation with energies of three subsystems (electric, magnetic, and magnetoelectric). The energy states of ME "trions" can be split in a bias magnetic field. For two opposite directions of a normal bias field, there are two types of ME "trions" with different chirality. Since the temporal inversion symmetry is broken, we can observe the right- and left-hand power-flow vortices [11]. Circulating energy contributes to the field's angular momentum. The combination of circulating energy (contributing to angular momentum) and the transfer of energy (propagation along an axis) results in a helical power flow. So, due to ME energy density we have *twisted wave propagation* of EM radiation in bianisotropic media.

The concept of ME quantum vacuum involves the study of vacuum states of quantized ME fields arising from the oscillation spectra of ME resonators. Fundamental studies of the interaction of a resonant point ME scatterer with EM radiation constitutes a field of research called ME quantum electrodynamics. The main aspects of the physics of ME meta-atoms should be related to magnetism, relativity theory, and topology. In near-field analysis, the role of ME energy is of great importance. The ME near fields are non-Maxwellian fields with space-time symmetry breaking. In the dynamic regime, the observation of ME characteristics in vacuum EM fluctuations is the subject of much discussion. Quantum fluctuations near the ME material with violation of the *PT* discrete symmetries will produce a sort of *PT* violating atmosphere [25]. It is argued that this atmosphere induces new kinds of "Casimir" forces on bodies near the material. There are claims that the vacuum can impart momentum asymmetrically on ME structures. Asymmetric momentum transfer arises from the ME structure since it breaks the temporal and spatial symmetries of electromagnetic modes. The possibility of extracting linear momentum from vacuum was discussed by Feigel [1]. Feigel argued that the momentum of vacuum zero fluctuations can occur only in a structure with *PT*-symmetry breakings. The main idea is to suggest a new quantum mechanical effect, namely the extraction of momentum from the electromagnetic vacuum oscillations. In the proposed effect, linear momentum is extracted from a vacuum field. This is different from the case of the Casimir effect, in which energy is extracted from a vacuum field [26]. It was argued that rotating ME particles can generate changes in momentum of zero-point fluctuations, which result in the "self-propulsion" in quantum vacuum. The "self-propulsion" in quantum vacuum requires mechanical back-action from ME particle. To provide this, the ME particle should be a propellor-like device [2]. Thus, the fundamental question of extracting angular momentum from the vacuum arises.

Finally, it is worth noting that while in "real" atoms the electric and magnetic energies (fields) are profoundly coupled due to relativistic effects, most noticeably manifested in the spin-orbit interaction, in ME meta-atoms the coupling of electric and magnetic energies (fields) occurs differently, but also due to relativistic effects and spin-orbit interaction [12].

## III. BIANISOTROPIC STRUCTURES BASED ON MAGNON-PLASMON META-ATOMS

In a quasi-2D ferrite disk particle with surface metal strip, electrical and magnetic properties are intrinsically cross-coupled. Such a magnon-plasmon meta-atom is shown on Fig. 1. For EM radiation, this ME meta-atom is viewed as a subwavelength scatterer with a multiresonant ME response spectrum [27 – 30]. The experimental results with the ferrite-based ME meta-atom, presented in Fig. 2, indicate the existence of quantized ME energy – the energy of interaction between quasistatic oscillations of the electric and magnetic subsystems [29]. The magnon-plasmon magnetoelectricity refers to a specialized phenomenon where magnetic excitations (magnons) and charge oscillations (plasmons) couple to create unique ME properties in a subwavelength region of EM radiation. The ME interaction is mediated by magnetic-dipolar-mode (MDM) oscillations and topology effects, which exhibit broken symmetries that allow for the co-existence of electric and magnetic moments. The physics of this macroscopic quantum effect of an energy conversion between the average stored magnetic, electric, and ME energies differs from the microscopic effect of magnon-plasmon hybridization, where the electric field associated with plasmon oscillations creates a nonequilibrium spin density that couples to magnons by an exchange interaction [31]. Fig. 3 illustrates fundamentally different concepts of two types of bianisotropic metamaterials. In a bianisotropic metamaterial composed by *LC*-circuit elements [32], we have ME responses due to far-field EM radiation. In metamaterials composed by magnon-plasmon meta-atoms, *local* ME responses are observed caused by internal (not only electromagnetic) dynamic processes.

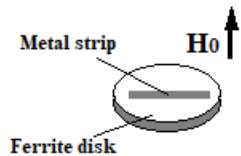

Fig. 1. A quasi-2D ferrite disk particle with surface metal strip.

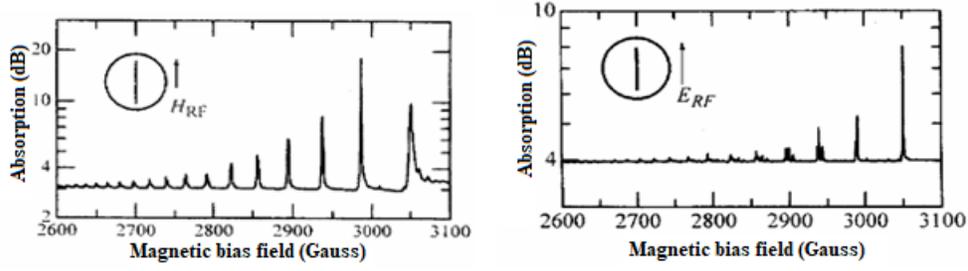

Fig. 2. Experimental results demonstrating the existence of quantized ME energy in the magnon-plasmon ME meta-atom. The complete coincidence of the spectral peaks of quasistatic oscillations of the electric and magnetic subsystems indicates the existence of ME oscillations with quantized energy levels.

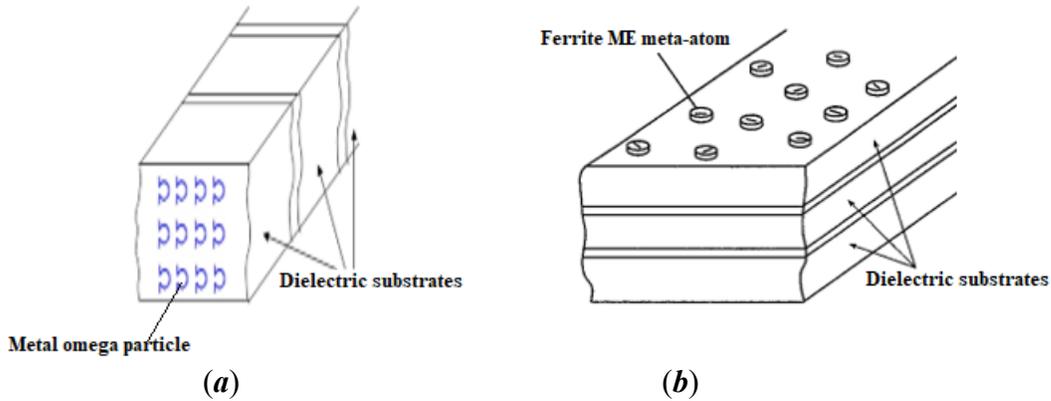

Fig. 3. Two fundamentally different concepts of ME metamaterials. (*a*) Bianisotropic materials with *LC*-circuit elements (omega particles). ME responses are observed as far-field characteristics. (*b*) ME meta materials composed of special ferrite resonators. Response with quantized ME energy can be observed in the near-field regions.

In a bianisotropic structure based on magnon-plasmon meta-atoms, each subwavelength particle may be considered as a glued pair of two magnetic and electric dipoles: the magnetic dipole is due to the ferrite body and the electric dipole is due to the metallization region. When such a particle is described quasistatically and is considered as a $\delta$-functional dipolar scatterer, one can use the integral-form constitutive relations (ICR) for a bianisotropic material:

$$D_i(t,\vec{r}) = (\varepsilon_{ij} \circ E_j) + (\xi_{ij} \circ H_j), \qquad (6)$$

$$B_i(t,\vec{r}) = (\zeta_{ij} \circ E_j) + (\mu_{ij} \circ H_j). \qquad (7)$$

The integral operators on the right-hand side of these expressions have a form similar to the integral operator:

$$(\varepsilon_{ij} \circ E_j) = \int_{-\infty}^{t} dt' \int d\vec{r}' \varepsilon_{ij}(t,\vec{r},t',\vec{r}') E_j(t',\vec{r}') . \tag{8}$$

The kernels of the operators in the above ICRs are responses of a medium to the $\delta$-function electric and magnetic fields. Convergence of integrals in the ICRs can be proven if one shows a physical mechanism of influence of short-time and short-space *quasistatic* interactions on the medium polarization properties.

In an assumption that ME material strictures based on ferrite magnon-plasmon meta-atom can be realized, theory of bianisotropic crystal lattices, has been proposed [33]. To use macroscopic Maxwell's equations for the material continuum, the maximum scale of material nonhomogeneity must be much less than distances of macroscopic field variations. For electromagnetic waves these distances correspond to the wavelength. Because of the limiting cutoff wave numbers, all variables in macroscopic electrodynamics are finite-spectrum functions [34]. Two ways may be used to describe the electromagnetic field-condensed media interaction. When one way is to get over a discrete structure of a medium by the averaging procedure, another way may be conceived as follows: to *discretize fields* based on the discrete structure of a medium. When initial restrictions to the wave number spectrum take place, one can use the so-called *sampling theorem* for a medium modeled as a triple infinite periodic array of identical $\delta$-functional scattering elements. Taking into account the Lorenz-Lorentz model one can develop a dynamical theory which considers strong field fluctuations in crystal lattices [35]. The method used in [35] also becomes important for the dynamical model of bianisotropic crystal lattices when every bianisotropic particle is a $\delta$-functional dipolar scatterer of coupled electric and magnetic dipoles [33].

However, by using the results of the sampling theorem for a crystal lattice composed by $\delta$-functional scattering ME elements, we effectively ignore the conclusion made in Section II that electromagnetic radiation propagates in the form of twisted waves in bianisotropic media. To clarify this situation, let us consider the near-field structure of our magnon-plasmonic meta-atoms. Figure 4 shows the electric field distribution on the surface of a metal strip for different time phases at a given direction of the bias magnetic field. A clear asymmetry shift is observed in this distribution of a field. With the bias magnetic field in the opposite direction, the direction of the asymmetry shift reverses.

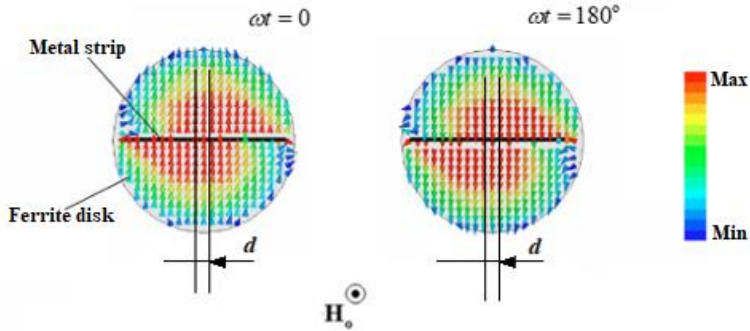

Fig. 4. The electric field distribution on the surface of a metal strip for different time phases at a given direction of the bias magnetic field. The direction of the asymmetry shift *d* is reversed when the bias magnetic field is in the opposite direction.

The asymmetry of the presented distribution is due to the orbital rotation of the near-field structure. In a magnetized ferrite disk, the electromagnetic Goos–Hänchen shift is closely linked to the generation of *orbital angular momentum and chiral states* due to the breaking of time-reversal symmetry. When an EM wave interacts with a magnetized ferrite, the non-reciprocity induced by the external magnetic field causes a lateral Goos–Hänchen shift. In small resonant structures like a disk, this lateral displacement – combined with the internal reflections – manifests itself as a rotational motion of the energy flow. This symmetry breaking results in the formation of EM vortices. Depending on the direction of magnetization, these resonances exhibit opposite vortex rotations (different chirality).

The above aspects indicate that a subwavelength ferrite disk by itself – without a metal strip on its surface – can behave as a ME particle with *quantized orbital angular momentum*. Our study suggests the existence of such ME resonances [36]. These resonances involve coupled states of magnetostatic (MS) and electrostatic (ES) functions within the subwavelength domain. The ME resonances are intrinsically linked to ME energy. The concept of "quantum resonances" relates to the quantization of internal magnetic energy or magnetization states within the confined ferrite structure. The sharp, "atomic-like" spectra observed in microwave experiments with ferrite disks are schematic representations of these discrete states.

## III. ME QUANTUM RESONANCES IN A SUBWAVELENGTH FERRITE DISK

While the understanding of magnetism (and ultimately electromagnetism) arose from electricity, our understanding of magnetoelectricity arose from magnetism. Violation of the invariances under time inversion *T* and space reflection parity *P* in the magnetic structure are necessary conditions for the emergence of the ME effect. In the expression for free energy, the energy term describing the ME coupling has the form:

$$W_{ME} = \alpha_{ij} E_i H_j, \tag{9}$$

Where $\alpha_{ij}$ is a ME tensor [6]. For linear ME effect, both *T* and *P* should be broken, but the product *PT* is conserved. The magnitude of the magnetoelectric coefficient in a medium is ultimately restricted by the speed of light in that medium.

Dynamic magnetoelectricity in a ME meta-atom is due to generation of electrical polarization by means of magnetization changing in space and time. These are dynamic properties of resonantly oscillating magnetizations and *magnetically induced* polarizations. Electric polarization can be induced due to a topological effect associated with *chiral magnetic currents* at the boundary of a sample. These are topologically protective edge currents in a *non-simply connected domain*. While analyzing what is a suitable geometrical shape of a 3D confined subwavelength ferrite sample where topological edge currents can occur, we should not consider the spherical sample because the sphere is simply connected and thus every current loop can be contracted on the surface to a point. Samples should also not be in the shape of an ellipsoid and an infinite cylinder. At the same time, a quasi-2D disk turns out to be the suitable sample shape. When the current loop is on the lateral surface of such a disk, we have a non-simply connected domain with topologically protective edge currents.

In a sample made of a magnetic insulator, an interaction between the ferromagnetic order subsystem and the electric polarization subsystem can occur. Similar to type II multiferroics, the intrinsic ME coupling in the ferrite disk is due to the electric polarization caused by spatially and temporally modulated spin structures, but the physical mechanism is completely different. In ME meta-atoms, the mesoscopic effect of dynamic magnetoelectricity arises from the topologically coupled MS and ES resonances. Both MS and ES oscillations are observed with quantum confinement effects for scalar *wave functions* $\psi(\vec{r},t)$ and $\phi(\vec{r},t)$ in a small sample localized in the subwavelength region of EM radiation. These wave functions are introduced as quasistatic solutions of the magnetic and electric fields: $\vec{H} = -\vec{\nabla}\psi$ and $\vec{E} = -\vec{\nabla}\phi$. The term "ME coupled" implies that the MS and ES states are not isolated but rather interact with each other due to topologically induced edge currents. In general, this interaction is expressed by terms in the Hamiltonian that relates to the ME states: $\langle\psi|\mathrm{H}_{ME}|\phi\rangle$. The ME coupled states can be represented using a tensor product of the individual, MS and ES, state spaces. In the ME resonance states, a violation of both spatial and temporal inversion symmetry occurs. ME duality involves a symmetry relationship between time-varying electric and magnetic fields that is distinct from EM fields. Any EM retardation effects are disregarded.

Due to circulation of a chiral magnetic current on a lateral surface of a ferrite disk, electric charges appear on the top and bottom planes of the ferrite disk. This induces a normal electric-field gradient. As a result, we have the *electric-quadrupole precession caused by the magnetization dynamics*. The magnetic-dipole and electric-quadrupole resonances are well known in nuclear physics [37]. The electric field gradient arises from the inhomogeneous distribution of charges. Electric quadrupole precession, in the context of nuclear magnetic resonance (NMR) refers to the precession of nucleus's quadrupole moment in an electric field gradient. This precession is distinct from regular Larmor precession, which is caused by a magnetic field. In our case, we have a unique effect of coupling these resonances. The spectral analysis for ES scalar wave functions $\phi$ in a quasi-2D ferrite disk will be like the spectral analysis for MS scalar wave functions $\psi$. Similar to magnetic oscillations, for electric oscillations we have *electric chiral currents* at the lateral boundary of the disk sample. Because of this topologically protective electric edge currents the magnetic-field flux is observed.

The model predicts the topological ME effect, where an edge current of orbital magnetization generates a topological contribution to electric polarization and an edge current of orbital polarization generates a topological contribution to magnetization. Topological currents provide us with the possibility to have chiral rotational symmetry by the turn over a regular-coordinate angle $2\pi$ at the $\pi$-shift of a dynamic phase of the external EM field. The frequency of orbital rotation must be twice the EM wave frequency $\omega$. In the coordinate frame of orbitally driven field patterns, the lines of the electric field $\vec{E}$ as well the lines of the polarization $\vec{p}$ are "frozen" in the lines of magnetization $\vec{m}$. It means that there are no time variations of vectors $\vec{E}$ and $\vec{p}$ with respect to vector $\vec{m}$.

The physics of quantum ME resonances in a subwavelength ferrite particle is based on an understanding of the main aspects of the theory of spectral properties of MDMs in a ferrite disk resonator, published in [30, 38 – 44]. Here we present brief excerpts from this theory.

When analyzing MDM oscillations in a ferrite disk, two types of solutions for scalar wave function $\psi(\vec{r},t)$ are considered. There are the spectral solutions of the energy eigenstates, conventionally called the *G*-modes solutions, and the spectral solutions of the power-flow-confinement states, conventionally called the *L*-modes solutions [30, 38 – 44]. For *G* modes, we define the energy eigenstates of MS oscillations based on the Schrödinger-like equation for scalar wave function $\psi(\vec{r},t)$ with use of the Neumann-Dirichlet (ND) boundary conditions. In case of *L* modes, we consider normalization to the power-flow density

$$\vec{\mathcal{J}} = \frac{i\omega}{4}\left(\psi \vec{B}^* - \psi^* \vec{B}\right) \tag{10}$$

using the EM boundary conditions. Obviously, when characterizing the MDM oscillations, the resonant states of the *G* and *L* modes should be considered together.

Analyzing the *G*-mode solutions in a cylindrical coordinate system $(z, r, \theta)$, we determine the membrane function $\tilde{\eta}(r,\theta)$ by the Bessel-function order and the number of zeros of the Bessel function corresponding to the radial variations. Membrane functions $\tilde{\eta}(r,\theta)$ is a single-valued function. On a lateral surface of a ferrite disk, the ND boundary conditions for mode *n* are written as

$$\left(\tilde{\eta}_n\right)_{r=\mathcal{R}^-} - \left(\tilde{\eta}_n\right)_{r=\mathcal{R}^+} = 0 \tag{11}$$

and

$$\mu \left(\frac{\partial \tilde{\eta}_n}{\partial r}\right)_{r=\mathcal{R}^-} - \left(\frac{\partial \tilde{\eta}_n}{\partial r}\right)_{r=\mathcal{R}^+} = 0, \tag{12}$$

where $\mathcal{R}$ is a disk radius. For *G* modes, the spectral problem gives the energy orthogonality relation: $(E_n - E_{n'}) \int_{S_c} \tilde{\eta}_n \tilde{\eta}_{n'}^* dS = 0$. The quantity $E_n$ is considered as density of accumulated magnetic energy of mode *n*. This is the average (on the RF period) energy accumulated in a flat ferrite-disk region of in-plane cross-section and unit length along *z* axis. Since the space of square integrable functions is a Hilbert space with a well-defined scalar product, we can introduce a basis set. The mode amplitude can be interpreted as the probability to find a system in a certain state *n*. Using the principle of wave-particle duality, one can describe this oscillating system as a collective motion of quasiparticles. There are "flat-mode" quasiparticles at a reflexively-translational motion behavior between the lower and upper planes of a quasi-2D disk. Such quasiparticles are called "light" magnons. In our study we consider *"light" magnons in ferromagnet as quanta of collective MS spin waves* that involves the precession of many spins on the long-range dipole-dipole interactions. It is different from the short-range magnons for exchange-interaction spin waves with a quadratic character of dispersion. The meaning of the term "light", used for the condensed MDM magnons, arises from the fact that effective masses of these quasiparticles are much less, than effective masses of "real" magnons – the quasiparticles describing small-scale exchange-interaction effects in magnetic structures. The *effective mass* of the "light" magnon for a monochromatic MDM is defined as [40]:

$$\left(m_{lm}^{(eff)}\right)_n = \frac{\hbar}{2}\frac{\beta_n^2}{\omega}, \tag{13}$$

where $\beta_n$ is the propagation constant of mode $n$ along the disk axis $z$.

In the *L*-mode solutions, we determine the membrane function $\tilde{\varphi}(r,\theta)$ at the same way as the membrane function $\tilde{\eta}(r,\theta)$ for *G*-mode solutions. The continuity of $\tilde{\varphi}(r,\theta)$ on a lateral surface of a ferrite disk is characterized by the equation like Eq. (11). However, for the derivatives on a lateral surface we have nonhomogeneous boundary conditions:

$$\mu\left(\frac{\partial \tilde{\varphi}_n}{\partial r}\right)_{r=\mathcal{R}^-} - \left(\frac{\partial \tilde{\varphi}_n}{\partial r}\right)_{r=\mathcal{R}^+} = -\left(i\mu_a \frac{1}{r}\frac{\partial \tilde{\varphi}_n}{\partial \theta}\right)_{r=\mathcal{R}^-}. \tag{14}$$

This is the EM boundary condition of continuity of the magnetic flux density on a lateral surface of a ferrite disk. In Eqs. (12) and (14), $\mu$ and $\mu_a$ are diagonal component and off-diagonal components of the permeability tensor $\vec{\mu}$ [45]. When using the EM boundary conditions, it becomes obvious that the membrane function $\tilde{\varphi}(r,\theta)$ must not only be continuous and differentiable with respect to a normal to the lateral surface of the disk, but, because of the presence of a gyrotropy term, be also differentiable with respect to a tangent to this surface.

From Eq. (14), it follows that for a given direction of a bias magnetic field (which defines a sign of $\mu_a$), we observe both clockwise (CW) and counterclockwise (CCW) azimuthally propagating modes. In this case, it can be assumed that the strength of the scattering of light for the clockwise to the counterclockwise propagation direction is different. For homogeneous ND boundary condition, we have Hermitian Hamiltonian resulting in real energy eigen states. However, since the sample is an open system, the coupling to the environment expressed by the EM boundary conditions leads to non-Hermitian Hamiltonian with complex wave functions. Membrane functions $\tilde{\varphi}(r,\theta)$ are not single-valued functions. It can be represented as a two-component spinor [44]:

$$\tilde{\varphi}_n(\vec{r},\theta) = \tilde{\eta}_n(\vec{r},\theta)\begin{bmatrix} e^{-\frac{1}{2}i\theta} \\ e^{+\frac{1}{2}i\theta} \end{bmatrix} \tag{15}$$

Circulation of gradient $\vec{\nabla}_\theta \tilde{\varphi}$ along contour $L = 2\pi r$ is not equal to zero. On a lateral border of a ferrite disk ($r = \mathcal{R}$), we express function $\tilde{\varphi}$ as

$$\tilde{\varphi} = \tilde{\eta} \delta_\pm, \qquad (16)$$

Where $\delta_\pm$ is a double-valued edge wave function on contour $\mathcal{L} = 2\pi\mathcal{R}$ [38, 41].

On a lateral surface of a quasi-2D ferrite disk, one can distinguish two different functions $\delta_\pm$, which are the counterclockwise and clockwise rotating-wave edge functions with respect to a membrane function $\tilde{\eta}(r,\theta)$. The spin-half wave-function $\delta_\pm$ changes its sign when the regular-coordinate angle $\theta$ is rotated by $2\pi$. As a result, one has the eigenstate spectrum of MDM oscillations with topological phases accumulated by the edge wave function $\delta$. A circulation of gradient $\vec{\nabla}_\theta \delta = \frac{1}{r}\left(\frac{\partial \delta_\pm}{\partial \theta}\right)_{r=\mathcal{R}} \vec{e}_\theta$ along contour $\mathcal{L} = 2\pi\mathcal{R}$ gives a non-zero quantity when an azimuth number is $q_\pm = \pm\frac{1}{2}, \pm\frac{3}{2}, \pm\frac{5}{2}...$ A line integral around a singular contour $\mathcal{L}$: $\frac{1}{\mathfrak{R}} \oint_\mathcal{L} \left(i\frac{\partial \delta_\pm}{\partial \theta}\right)(\delta_\pm)^* d\mathcal{L} = \int_0^{2\pi} \left[\left(i\frac{\partial \delta_\pm}{\partial \theta}\right)(\delta_\pm)^*\right]_{r=\mathfrak{R}} d\theta$ is an observable quantity. It follows from the fact that because of such a quantity one can restore single-valuedness (and, therefore, Hermicity) of the spectral problem. We can represent this observable quantity as a linear integral of a certain vector potential. Because of the existing the geometrical phase factor on a lateral boundary of a ferrite disk, MDM oscillations are characterized by a pseudo-electric field (the gauge field) $\vec{\epsilon}$. The pseudo-electric field $\vec{\epsilon}$ can be found as $\vec{\epsilon}_\pm = -\vec{\nabla} \times (\vec{\Lambda}_\epsilon^{(m)})_\pm$, where a vector function $(\vec{\Lambda}_\epsilon^{(m)})_\pm$ can be considered as the Berry connection. The gauge-invariant field $\vec{\epsilon}$ is the Berry curvature. The corresponding flux of the field $\vec{\epsilon}$ through a circle of radius $\mathfrak{R}$ is obtained as: $K \int_S (\vec{\epsilon})_\pm \cdot d\vec{S} = K \oint_\mathcal{L} (\vec{\Lambda}_\epsilon^{(m)})_\pm \cdot d\vec{\mathcal{L}} = K(\Xi^{(e)})_\pm = 2\pi q_\pm$, where $(\Xi^{(e)})_\pm$ are quantized fluxes of pseudo-electric fields and $K$ is the normalization coefficient. The physical meaning of coefficient $K$ concerns the property of a flux of a pseudo-electric field. There are the positive and negative eigenfluxes. In the MDM spectral problem, it is impossible to satisfy the EM boundary conditions without a flux $(\Xi^{(e)})_\pm$. Each MDM is characterized by energy eigenstate and is quantized to a quantum of an emergent electric flux.

In our system, there should be a certain internal mechanism which creates a nonzero vector potential $(\vec{\Lambda}_\epsilon^{(m)})_\pm$. This internal mechanism becomes apparent when comparing the ND boundary condition (12) (providing single-valuedness) and the EM boundary condition (14) (which does not provide single-valuedness). The difference arises from the term in the right-hand side in Eq. (14), which contains the gyrotropy parameter, the off-diagonal component of the permeability tensor $\mu_a$, and the annular magnetic field $\vec{H}_\theta = -\frac{1}{r}\left(\frac{\partial \delta_\pm}{\partial \theta}\right)_{r=\mathcal{R}} \vec{e}_\theta$. Just due to this term a nonzero vector potential appears. The annular magnetic field $\vec{H}_\theta$ is a singular field existing only in an infinitesimally narrow cylindrical layer abutting from a ferrite side to a border of a ferrite disk. One does not have any special conditions connecting radial and azimuth components of magnetic fields on other inner or outer circular contours, except contour $\mathcal{L} = 2\pi\mathcal{R}$. Because of such an annular magnetic field, the notion of an effective circular magnetic current can be considered. The Berry mechanism provides a basis for the surface magnetic current at the interface between gyrotropic and nongyrotropic media. Following the spectrum analysis of MDMs in a quasi-2D ferrite disk one obtains edge chiral magnetic currents. This results in appearance of an anapole moment. For mode $n$, the anapole moment is calculated as [38, 41].

$$\left(a_\pm^{(e)}\right)_n \propto \mathcal{R} \int_0^d \oint_\mathcal{L} \left[\vec{j}_s^{(m)}(z)\right]_n \cdot \vec{dl}\, dz, \tag{17}$$

where $\vec{j}_s^{(m)}$ is the edge persistent magnetic current. At a large distance from the disk, localized distribution of an edge magnetic current is viewed as an electric field $\vec{a}^{(e)}$. The electric moment $\vec{a}^{(e)}$ is considered as the density of the electric flux $\Xi^{(e)}$.

Based on Eq. (10) for the power-flow density, in Ref. [41] it was shown that the orthogonality conditions for the $L$-mode spectral solutions take place when

$$\int_0^{2\pi} \left(\mathcal{T}_\pm^{(s)}(z)\right)_\theta d\theta = \frac{1}{4}\omega\mu_0 \Re \int_0^{2\pi} \left[\left(i\mu_a \frac{\partial \delta_\pm}{\partial \theta}\right)^* \delta_\pm - \left(i\mu_a \frac{\partial \delta_\pm}{\partial \theta}\right)(\delta_\pm)^*\right]_{r=\Re} d\theta = 0 \tag{18}$$

where $\left(\mathcal{T}_\pm^{(s)}\right)_\theta$ is the surface power-flow density on contour $\mathcal{L} = 2\pi\mathcal{R}$. It is evident, however, that the power-flow-confinement states can be realized when a softer boundary condition on contour $\mathcal{L}$ is used:

$$\vec{\nabla}_\theta \cdot \left[\left(\mathcal{T}_\pm^{(s)}(z)\right)_\theta\right]\vec{e}_\theta = \frac{1}{\Re}\frac{\partial}{\partial \theta}\left[\left(i\mu_a \frac{\partial \delta_\pm}{\partial \theta}\right)^* \delta_\pm - \left(i\mu_a \frac{\partial \delta_\pm}{\partial \theta}\right)(\delta_\pm)^*\right]_{r=\Re} = 0. \tag{19}$$

This implies the presence of the edge persistent power-flow circulation.

Due to the surface power flow density, the membrane eigenfunction $\tilde{\eta}$ of the MDM rotates around the disk axis. When for every MDM we introduce the notion of an effective mass $\left(m_{lm}^{(eff)}\right)_n$, expressed by Eq. (13), we can assume that for every MDM there exists also an *effective moment of inertia* $\left(I_z^{(eff)}\right)_n$. With this assumption, an orbital angular momentum a mode is expressed as $(L_z)_n = (I_z^{(eff)})_n \omega$. Supposing, as the first approximation, that the membrane eigenfunction $\tilde{\eta}_n$ is viewed as an infinitely thin homogenous disk of radius $\mathcal{R}$ ( in other words, assuming that for every MDM, the radial and azimuth variation of the MS-potential function, are averaged), we can write for the effective moment of inertia

$$\left(I_z^{(eff)}\right)_n = \frac{1}{2}\left(m_{lm}^{(eff)}\right)_n \mathcal{R}^2 d \ . \tag{20}$$

The orbital angular momentum a mode is expressed as

$$\left(L_z^{(eff)}\right)_n = \left(I_z^{(eff)}\right)_n \omega = \frac{\hbar}{4}\beta_n^2 \mathcal{R}^2 d \ . \tag{21}$$

With use of the EM boundary conditions, we consider the spectral solutions for the MS wave functions $\psi$ as generating functions for determining the fields. For any mode *n*, magnetization field is found as $\vec{m} = -\frac{1}{4\pi}(\vec{\mu} - \vec{I})\vec{\nabla}\psi$ [45]. Knowing $\vec{\nabla}\cdot\vec{m}$ and $\vec{\nabla}\times\vec{m}$, we can obtain for the electric and magnetic fields outside a ferrite disk [38]:

$$\vec{E}(\vec{r}) = -\frac{i\omega\mu_0}{4\pi}\left(\int_V \frac{\vec{\nabla}'\times\vec{m}(\vec{r}')}{|\vec{r}-\vec{r}'|}dV' + \int_S \frac{\vec{m}(\vec{r}')\times\vec{n}'}{|\vec{r}-\vec{r}'|}dS'\right) \tag{22}$$

and

$$\vec{H}(\vec{r}) = \frac{1}{4\pi}\left(\int_V \frac{(\vec{\nabla}'\cdot\vec{m}(\vec{r}'))(\vec{r}-\vec{r}')}{|\vec{r}-\vec{r}'|^3}dV' - \int_S \frac{(\vec{n}'\cdot\vec{m}(\vec{r}'))(\vec{r}-\vec{r}')}{|\vec{r}-\vec{r}'|^3}dS'\right) , \tag{23}$$

where *V* and *S* are a volume and a surface of a ferrite sample, respectively. Vector $\vec{n}'$ is the outwardly directed normal to surface *S*.

In the vacuum near-field region adjacent to the MDM ferrite disk, there exist power-flow vortices, defined by the cross product $\text{Re}(\vec{E}\times\vec{H}^*)$. Together with this, there is another quadratic-form parameter determined by a *scalar product* between the electric and magnetic field components [38, 44]:

$$F = \frac{\varepsilon_0}{4}\text{Im}\left[\vec{E}\cdot(\nabla\times\vec{E})^*\right] = \frac{\omega\varepsilon_0\mu_0}{4}\text{Re}(\vec{E}\cdot\vec{H}^*) = \frac{\omega\varepsilon_0\mu_0}{4}\text{Re}(\vec{\nabla}\phi\cdot\vec{\nabla}\psi^*) \tag{24}$$

The presence of this parameter, called the parameter of helicity, in the vacuum region is a fundamental effect in our analysis. In Ref. [46, 47], in was argued, that a linear structure of the EM radiation fields in vacuum can be observed only when $\vec{E}\cdot\vec{B}=0$. At this condition, the electromagnetic helicity is defined as a difference between the numbers of right- and left-handed photons. On the other hand, when $\vec{E}\cdot\vec{B}\neq 0$, quantum electrodynamics predicts that the vacuum behaves like a material medium. In this case, the linear Maxwell theory receives nonlinear corrections. One can observe such a ME birefringence of the quantum vacuum when static magnetic and electric fields are applied [48 – 50]. In our case, the nonlinearity in the vacuum near-field region adjacent to the MDM ferrite disk arises due to magnon-magnon dipole interaction effects. We observe the dominant magnonic response, even at room temperature. Large binding energy and small size of a MDM particle enables strong light-matter coupling to cavity photons and magnons, leading to emergent magnon-polaritons.

The helicity parameter in (24) actually represents the *ME energy density*. In Ref. [42], we have qualitatively explained how the multiresonance states in a microwave resonator, experimentally observed in [36, 51, 52], are associated with a quantized change in the energy of a ferrite disk, which arises due to an external source – a bias magnetic field $H_0$ – at a constant frequency of the microwave signal. It was stated that there is a quantum effect of electromagnetically generated demagnetization of a sample:

$$\Delta W^{(n)} = -\frac{1}{2}\int_V \vec{H}_0^{(n)} \cdot \Delta\left(\vec{M}^{(n)}\right)_{eff} dV \qquad (25)$$

The energy $\Delta W^{(n)}$ is the microwave energy extracted from the magnetic energy of a ferrite disk at the *n*-th MDM resonance. It was supposed that the demagnetizing magnetic field is reduced due to effective magnetic charges on a ferrite-disk planes. For the reduced DC magnetization of a ferrite disk, we have the frequency $\left(\omega_M^{(n)}\right)_{eff} = \gamma\mu_0 \left(M_0^{(n)}\right)_{eff}$, that is less than such a frequency $\omega_M = \gamma\mu_0 M_0$ in an unbounded magnetically saturated ferrite [45]. The quantized magnetic charges on the ferrite-disk planes are caused by the induced electric gyrotropy and orbitally driven electric polarization inside a ferrite. This is due to EDM oscillations in the electric subsystem.

Several illustrations related to spectral properties of ME oscillations in a ferrite disk resonator, are shown in Figs. 5 – 8.

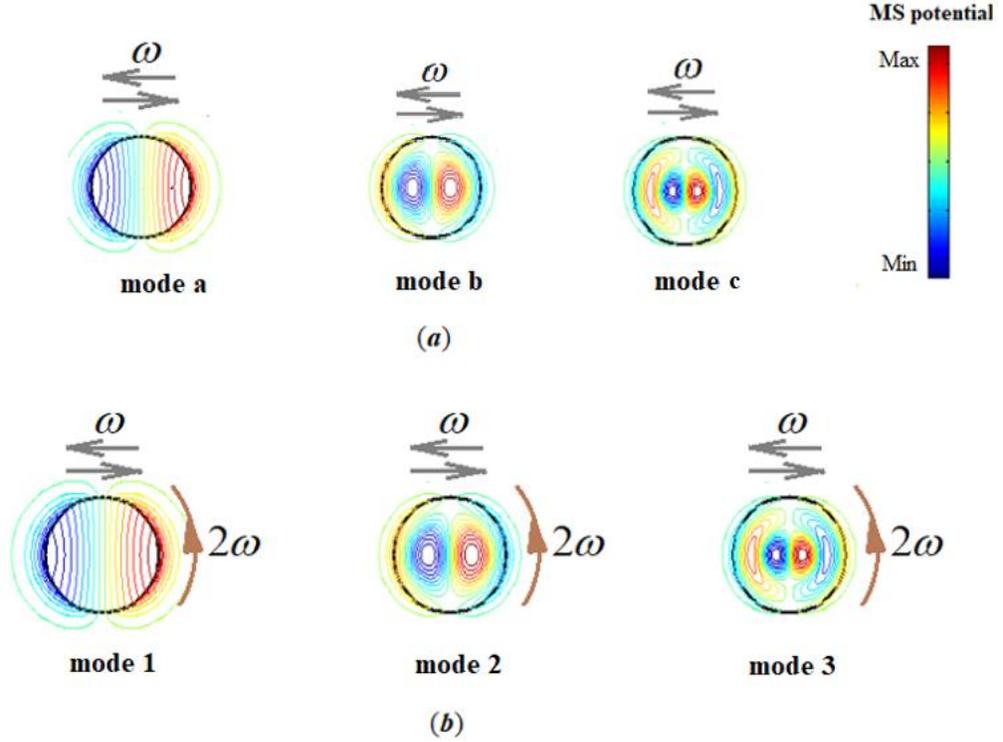

Fig. 5. MS-potential distributions for *G* and *L* modes in a quasi-2D ferrite disk viewed in the laboratory coordinate system. For *G* modes, there is a collinear magnetic system, where spins align in parallel or antiparallel configurations. For *L* modes, non-collinear arrangements exhibit spatially varying spin orientations that give rise to topologically non-trivial spin textures due to chiral rotation in systems lacking inversion symmetry. Lacking inversion symmetry is due to the involvement of the electric-dipole-polarization subsystem (see Fig. 6 below). (*a*) The first three *G* modes. (*b*) The first three *L* modes.

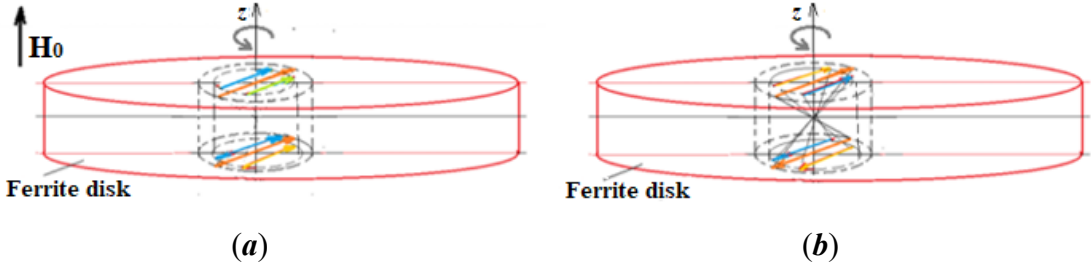

Fig. 6. The figure shows how local magnetic and electric moments inside the ferrite disk align and rotate in the *L* mode. In the laboratory coordinate system, we see magnetic-dipole (*a*) and electric-quadrupole (*b*) structures rotating at a frequency twice the frequence of the microwave signal. In a rotating coordinate system, the lines of polarization $\vec{p}$ are "frozen" in the lines of magnetization $\vec{m}$. The inversion symmetry is broken due to the electric-quadrupole structure (Ref. [53]). In magnetic resonance, the rotating frame simplifies the description of spin dynamics by effectively "freezing out" the time evolution of the spins due to the applied radiofrequency field.

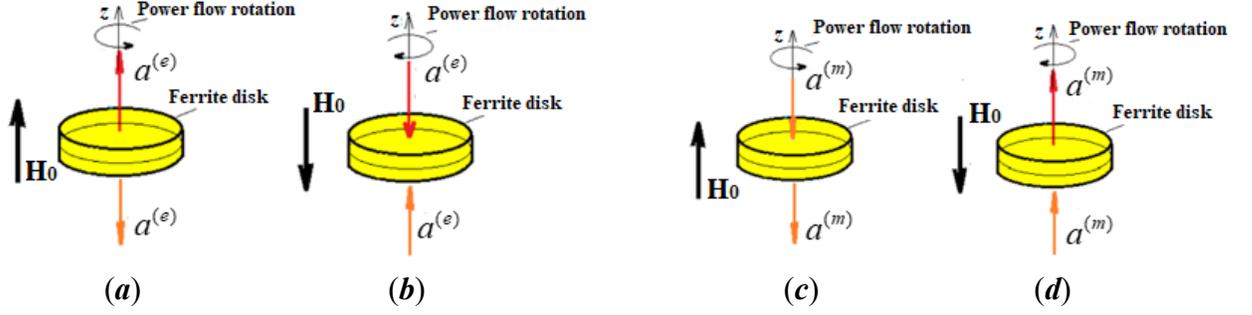

Fig. 7. Rotation may induce fictitious electric and magnetic charges. We observe electric and magnetic fluxes at the *L*-mode resonances. (***a***), (***b***) In the magnetic subsystem of the ferrite-disk resonator, localized distribution of an *edge magnetic current* is viewed as an *electric flux*. The electric moment $\vec{a}^{(e)}$ is considered as the density of the electric flux. This structure with opposite polarization directions on the top and bottom planes of a ferrite disk leads to appearance of the electric-field gradient along *z* axis. This electric-field gradient acts like a torque on the electric-quadrupole moment, causing it to precess. (***c***), (***d***) Localized distribution of an *edge electric current* in the electric subsystem is viewed as a dipole magnetic field at a large distance. The magnetic moment $\vec{a}^{(m)}$ is considered as the density of the *magnetic flux*. The arrows indicate the *z*-direction of the magnetic field flow at the center of the disk. (Ref. [53]).

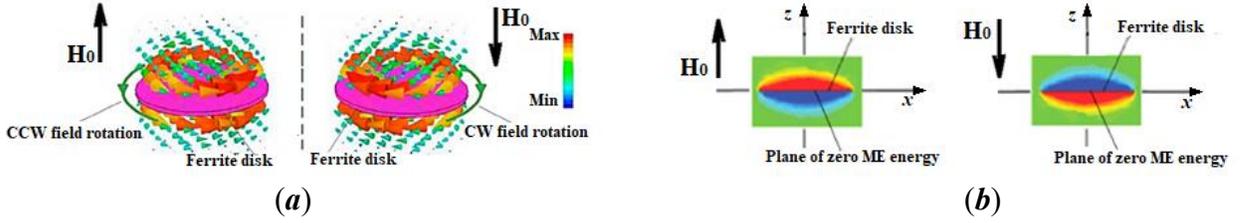

Fig. 8. Vacuum near-field region for the first *L* mode. (***a***) Power-flow rotation. (***b***) Positive (red colored) and negative (blue colored) ME energy (Ref. [44]).

Modes a, b, c shown in Fig. 5 are ground states. These are quasi-rest-frame structures. "Quasi-rest frame structures" refers to analyzing a system or object as if it were at rest, even though it might be in motion or undergoing acceleration, from the perspective of a non-inertial or accelerating frame of reference. This approach allows for simplification of complex dynamic problems by considering fictitious forces that arise due to the acceleration of the reference frame. Modes 1, 2, 3 in Fig. 5 are *metastable states*. *G* modes are MDMs. There is no ME coupling for *G* modes. But *L* modes are ME modes.

The main role of coupling of ME and ES resonance belongs to magnon *spin-orbit interaction* in a quasi-2D disk of magnetic insulators. In this case, we have a quasi-rotating ME system: A system that exhibits rotational characteristics (being analyzed within a rotating reference frame) but it's not a real rotation. In such a meta-atom, the effect of ME coupling vanishes in a rotating reference frame. The term "quasi-rotating" in this context refers to how the ME properties over time demonstrate a periodic pattern of changes.

## IV. HOW SUBWAVELENGTH MAGNETOELECTRIC PARTICLES INTERACT WITH CAVITY PHOTONS?

The spectral responses of a ferrite-disk ME meta-atom in a microwave waveguide and microwave cavity are defined by two external parameters – a bias magnetic field $H_0$ and a signal frequency $f$. The coherent quantum ME states are described with uncertainty in a bias magnetic field and frequency. When a magnetic field is suddenly changed, it introduces uncertainty in the energy of the magnetic system. Since energy and time are conjugate variables, a change in energy (due to the magnetic field change) will introduce an uncertainty in the time it takes for the system to adjust to the new field. This uncertainty in time corresponds to an uncertainty in the oscillation frequency. Since quantum transitions involve changes in energy levels, the uncertainty principle implies that the more precisely we know the energy change during a transition, the less precisely we can know the frequency. Due to the *uncertainty principle*, fluctuations in quantum fields, existing in a very narrow frequency deviation $\Delta f$ and very narrow region of a bias magnetic field $\Delta H_0$, can be considered as virtual particles. Beyond the frames of the uncertainty limit, one has a continuum of energy.

The spectra of ME oscillations in a quasi-2D ferrite disk are shown in Fig. 9. The sharply peaked (atomic-like) spectra from microwave experiments [36, 51, 52] may indicate macroscopic quantum phenomena with localized energy levels. In a waveguide cavity used in the aforementioned experiments, the frequency range of the entire MDM spectrum lies above the cut-off frequency of a dominant mode and below cut-off frequencies of the high-order modes of the cavity. It means that the complete-set spectrum of ME oscillations in a ferrite disk is viewed as virtual photons in the cavity.

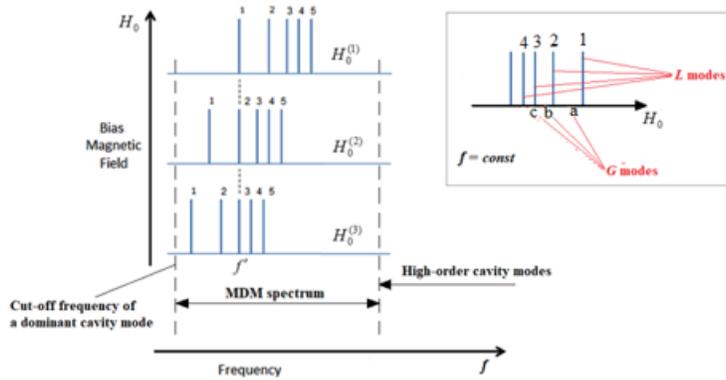

**Fig. 9.** Spectra of ME resonances of a ferrite disk in a cavity. For resonant mode positions, there is a clear correspondence between the frequency and the bias magnetic field. The entire ME spectrum lies in the frequency region where the high-order cavity modes are virtual photons. The inset shows the classification of the modes with respect to the bias magnetic field. $G$ modes are dark modes in a spectrum (absorption lines). $L$ modes are bright lines in a spectrum (emission lines). $G$ modes are MDMs, while $L$ modes are ME modes. The presented spectra show quantum transitions between $G$ and $L$ modes.

In the spectra shown on Fig. 9, one observes quantum transitions of $G$ and $L$ modes. This refers to the changes in the quantum state of a system, involving the transfer of energy and symmetry properties between these modes. The *G*-mode to *L*-mode transitions are dynamical *symmetry breaking transitions*. This describes the situation where the rotationally symmetric *G* mode undergoes a change in its fundamental (ground) state due to internal dynamics, leading to a less symmetrical (excited) *L*-mode state.

A point (local) ME particle shapes the structure of the EM field of the entire cavity space. In other words, a subwavelength ME meta-atom determines the field structure of the space-time. Figure 10 (**b**) shows the fields near a meta-atom in an excited ME state.

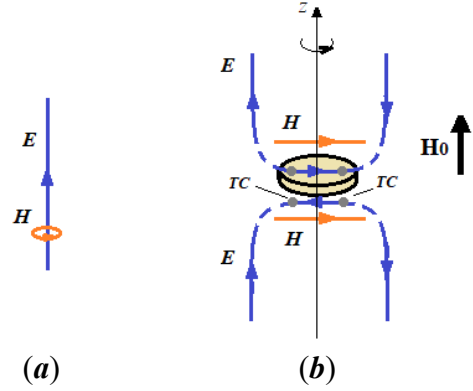

(*a*)          (*b*)

Fig. 10. The structure of the EM field. (*a*) Electric and magnetic fields in a microwave waveguide without a ferrite sample. (*b*) Field structure in the excited ME state near the meta-atom and on the upper and lower planes of the ferrite. TC denotes the topological charge.

In the case of a microwave cavity, we have two counter-propagating waves. Figure 11 shows the wavefront of the electric field in the cavity. We see curved wavefronts for counter-propagating waves in a cavity at a certain time phase and a given direction of a bias magnetic field. This is the structure of the electric field on a vacuum plane above the ferrite disk. A cavity polariton mode is a combination of the ME oscillations in ferrite disk with the EM radiation. The superposition of counter-propagating waves [images (a) and(b)] gives helical resonances of a polariton mode. Due to curved fronts, ME polaritons in a cavity have reduced energy.

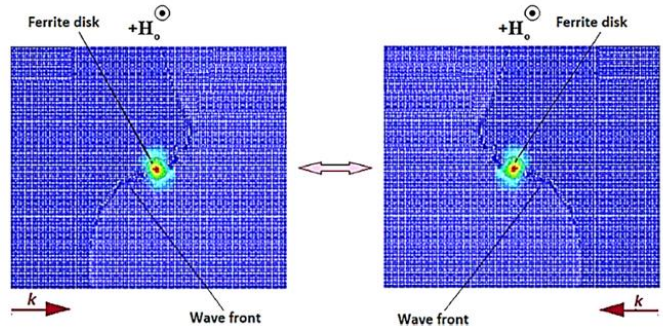

**Fig. 11.** ME polaritons in a cavity. Transformations of the wavefronts of the electric field for two counter-propagating EM waves occur at the excited ME states. *k* is the EM wavevector.

## V. ENTANGLEMENT BETWEEN TWO QUANTUM ME META-ATOMS

With curved wavefronts one obtains the spatial overlap of photons from separate ME meta-atoms. This is a necessary condition for high-quality entanglement generation. Entanglement between two quantum ME meta-atoms is a quantum phenomenon where two ferrite disks become intrinsically linked, so the quantum state of one instantly influences the state of the other. It can be mediated by a common microwave waveguide structure. Quantum entanglement between two ME quantum dots involves the creation of a unique quantum-mechanical coupling, in which the physical properties of the two ME dots and the ME photons they emit become strongly correlated, *regardless of the distance between them*. Excited local ME emitters interact due to the increased coherence length of quasistatic oscillations. A structure of two coupled ME meta-atoms embedded is a rectangular waveguide is shown in Fig. 12. A bias magnetic field has the same direction and the same quantity.

As a very attractive feature of the coupling, there is evidence for *long-range interaction* between two ME meta-atoms. This is the effect of tunneling of ME energy through EM waveguide channels. For ME oscillators, the effective EM wavelength stretches to infinity. ME meta-atoms interact and cooperate over a large spatial scale. While the correlation is instant, this phenomenon cannot be used to transmit information faster than light, as a classical channel is still required to interpret the measurement results. Figure 13 shows the position of wave fronts (localized regions where the electric field of the waveguide mode changes its sign) for coupled ferrite disks on a vacuum plane above the disks.

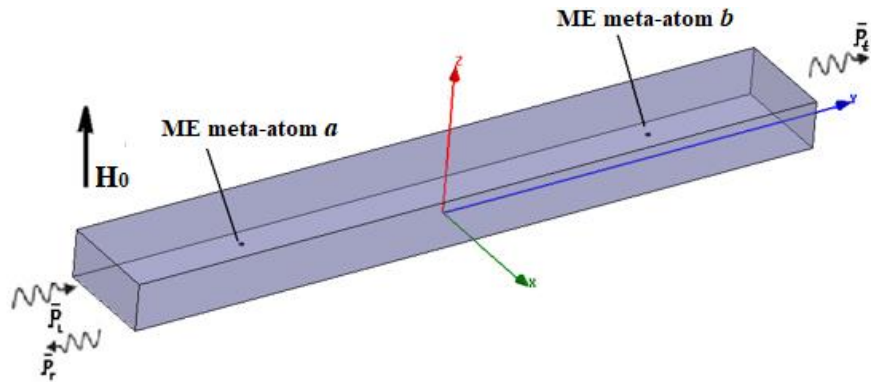

Fig. 12. A structure of two coupled ME meta-atoms embedded in a rectangular waveguide [54].

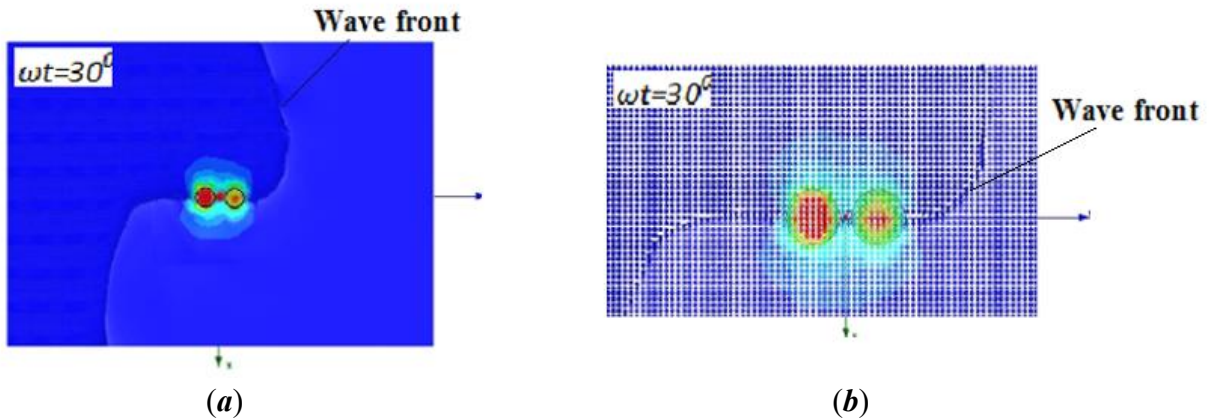

(*a*)          (*b*)

Fig. 13. (*a*) Positions of the wavefronts (localized regions where the electric field of a waveguide-mode changes its sign) for coupled ferrite disks on a vacuum plane above the disks. (*b*) enlarged pictures of the same distributions. The wavefronts are shown at a specific point in time. The positions of the wavefronts change over time.

Due to the coupling of ME meta-atoms in the microwave waveguide, a splitting of the ME resonance peak is observed. The frequency difference between the split peaks is very small. It is about 0.1% of the frequency of the ME resonance pick of a separate ME meta-atom. This splitting is shown in Fig. 14. Since the EM wavefront is located along the line connecting the disks (see Fig. 13), the effective wavelength of EM radiation along this line is extremely large. This is why the frequency difference between the split peaks is practically the same regardless of the distance between the meta-atoms.

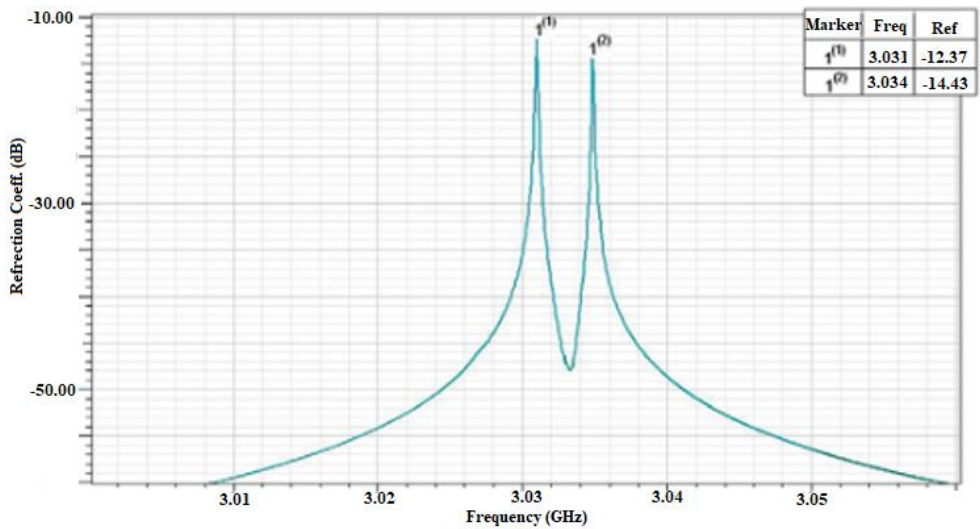

Fig. 14. Splitting of the ME resonance peak due to the coupling of ME meta-atoms in the microwave waveguide. The frequency difference between the split peaks is about 0.1% of the frequency of the ME resonance pick of a separate ME meta-atom. The frequency difference between the split peaks is practically the same regardless of the distance between the meta-atoms [54].

## VI. DISCUSSION AND CONCLUSION

In metamaterial structures, ME particles are considered as open resonators. To realize these structures as quantum systems with non-Hermitian Hamiltonian, the properties of *PT* symmetry must be taken into account. Obviously, *PT* symmetry is not applicable to enantiomeric chiral particles.

ME quantum resonances in a subwavelength ferrite disk give evidence to the fact that due to an external magnetic field (time-reversal symmetry breaking, *T*) and orbital chiral currents (space refraction symmetry breaking, *P*) these resonances are characterized by *biorthogonal* properties. Bianisotropic composites created from such ME particles will represent structures inherently described by the biorthogonal quantum mechanical formalism. Such *passive* systems with *PT* symmetry (having only losses, rather than balanced gain and loss) can be considered a more intriguing, or at least a more experimentally practical, area of research than *active* gain-loss systems, especially for quantum applications.

However, all of the above aspects appear insufficient, since the properties of such ME particles are not only closely related to the characteristics of the environment but also form the properties of that environment. In other words, quantized ME resonances in a ferrite disk can be said to create vacuum states of quantized ME fields. To clarify this, we have shown how subwavelength meta-atoms ME interact with a photon in a cavity. The interaction occurs through the *simultaneous breaking of time-reversal and space-inversion symmetries*, allowing for unique coupling between the electric and magnetic components of the fields in the cavity vacuum space. This is fundamentally different from conventional light-matter interactions and can lead to the formation of *novel hybrid light-matter states*. If a ME particle is under interaction with a "classical electrodynamics" object – the microwave cavity – the states of this classical object change. The character and value of these changes depend on the quantized states of the meta-atom and so can serve as its qualitative characteristics.

The concept of bianisotropy arose from the idea of creating materials that exhibit coupling between electric and magnetic fields on a local (subwavelength) scale. As we have discussed, in *quantum bianisotropy*, the light-matter interaction must be described by taking into account the *T* and *P* breaking symmetries in *both the ME particle and in the environment*. This means that we are talking about ME macroscopic quantum electrodynamics (ME-MQED). It is a theoretical framework that studies quantum light-matter interactions in a complex structure consisting of ME meta-atoms and ME environments. Here, ME environments means ME fields in vacuum (or dielectric), which possess both electric and magnetic components in a specific coupled way and break certain symmetries rather than being a simple combination of standard **E** and **H** fields. Both ME meta-atoms and ME environments are biorthognal systems.

Biorthogonality plays a central and essential role in the theoretical framework of *PT*-symmetry in non-Hermitian quantum mechanics. This is important in the theoretical understanding and practical realization of the ME effect, particularly regarding the balance of ME energy and power flow in metamaterials. In a composite structure consisting of subwavelength ME domains (meta-atoms), local *PT* symmetry may exist for an isolated component, but global PT symmetry breaking often occurs when these components interact with the environment such as a waveguide or cavity.